\documentclass[12pt]{article}
\pdfoutput=1     
  
\usepackage{hyperref}   
\usepackage{amsmath}      
\usepackage{amssymb} 
\usepackage{graphicx}
\usepackage{url}
\usepackage{enumerate}
\usepackage{color}
\usepackage{ulem}
 
\usepackage{float,wrapfig,color} 

\allowdisplaybreaks
\def\eq#1{(\ref{#1})}
\def\s[#1\s]{\begin{align}\begin{split}#1\end{split}\end{align}}
\def\[#1\]{\begin{align}#1\end{align}}

\def\bpsi{{\bar\psi}}
\def\bvphi{\bar \varphi}
\def\vphi{{\varphi}}
\def\pvphi{\varph_\parallel}
\def\pbpsi{\bpsi_\parallel}
\def\pvphi{\vphi_\parallel}
\def\ppsi{\psi_\parallel}
\def\pbvphi{\bvphi_\parallel}
\def\tbpsi{\bar \psi_\perp}
\def\tpsi{\psi_\perp}
\def\tvphi{\varphi_\perp}
\def\tbvphi{\bar \varphi_\perp}

\def\tphi{{\phi_\perp}}
\def\pphi{{\phi_\parallel}}
\def\tsig{{\sigma_\perp}}
\def\psig{{\sigma_\parallel}}

\begin{document}

\begin{titlepage} 

\title{
\hfill\parbox{4cm}{ \normalsize YITP-22-123}\\   
\vspace{1cm} 
Exact analytic expressions of real tensor eigenvalue \\ distributions
of Gaussian tensor model for small $N$ }

\author{Naoki Sasakura\footnote{sasakura@yukawa.kyoto-u.ac.jp}
\\
{\small{\it Yukawa Institute for Theoretical Physics, Kyoto University, }}\\
{\small {\it and } } \\
{\small{\it CGPQI, Yukawa Institute for Theoretical Physics, Kyoto University,}} \\
{\small{\it Kitashirakawa, Sakyo-ku, Kyoto 606-8502, Japan}}
}


\date{April 21, 2023}

\maketitle 

\begin{abstract}  
We obtain exact analytic expressions of real tensor eigenvalue/vector distributions 
of real symmetric order-three tensors with Gaussian distributions for $N\leq 8$.
This is achieved by explicitly computing the partition function of a zero-dimensional 
boson-fermion system with four-interactions. 
The distributions are expressed by combinations of polynomial, exponential and error functions
as results of feasible complicated bosonic integrals which appear after fermionic integrations.
By extrapolating the expressions and also using a previous result, we guess a large-$N$ expression.  
The expressions are compared with Monte Carlo simulations, 
and precise and good agreement are obtained with the exact and the large-$N$ expressions, respectively.
Understanding the feasibility of the integration is left for future study, which would
provide a general-$N$ analytic formula.
\end{abstract}
\end{titlepage}
 
\section{Introduction}
\label{sec:introduction}
Eigenvalue distributions are important dynamical quantities in studies of matrix models. 
They model energy eigenvalue distributions of complex dynamical systems \cite{Wigner}.
They provide a major technique in solving matrix models \cite{Brezin:1977sv}. 
Their topological properties differentiate phases of matrix models\footnote{For more details, 
see for instance \cite{matrix}.},
providing insights into the dynamics of gauge theories \cite{Gross:1980he,Wadia:1980cp}.

Recently, tensor models \cite{Ambjorn:1990ge,Sasakura:1990fs,Godfrey:1990dt,Gurau:2009tw} 
attract much attention in various contexts \cite{Ouerfelli:2022rus}.  
While it is important to develop efficient techniques of computing eigenvalues/vectors for certain tensors
in various practical applications of tensors \cite{qibook}, 
it is also interesting to study their distributions for ensembles of 
tensors, because tensors are dynamical in tensor models.

{{\color{blue}} While there are already some interesting results 
\cite{realnum1,realnum2,Evnin:2020ddw,Gurau:2020ehg, Sasakura:2022zwc,Sasakura:2022iqd}, 
eigenvalue distributions in tensor models still remain largely unexplored. 
In our previous studies \cite{Sasakura:2022zwc,Sasakura:2022iqd}, 
the problem was rewritten as computations of partition functions of zero-dimensional fermion systems.}
In \cite{Sasakura:2022zwc}, an exact formula of signed distributions of real eigenvalue/vector
distributions for real symmetric order-three tensors with Gaussian distributions 
was obtained, where each eigenvalue/vector contributed to the distribution by 
$\pm 1$, depending on the sign of a Hessian matrix associated to each eigenvalue/vector. 
In \cite{Sasakura:2022iqd},  real eigenvalue/vector distributions for the same ensemble
of tensors was studied. In this study, however, 
only an approximate large-$N$ expression\footnote{$N$ denotes 
the dimension of the index vector space of the tensor.} was obtained by truncating a Schwinger-Dyson equation,
though some closely related exact results were also obtained.
While the functional form of the large-$N$ expression, which was a Gaussian,  
agreed well with Monte Carlo simulations,
the overall factor did not and an improvement was expected.

In this paper, we rewrite the real eigenvalue/vector problem 
as a computation of a partition function of a boson-fermion system
with four-interactions, instead of the fermion systems in the previous studies
\cite{Sasakura:2022zwc,Sasakura:2022iqd}.
We exactly compute the real eigenvalue/vector distributions
for real symmetric order-three tensors with Gaussian distributions for $N\leq 8$,
and extrapolate the expressions to guess a large-$N$ expression, including the overall 
factor, which was not correctly obtained by the approximation of the previous paper \cite{Sasakura:2022iqd}.
We compare the exact expressions for small $N$ with Monte Carlo simulations, and obtain precise
agreement. We also compare the large-$N$ expression, and obtain good agreement. 

{{\color{blue}}
Computing the real eigenvalue/vector
distributions is essentially the same as the computation of complexity in the $p$-spin spherical model of spin 
glasses \cite{pspin,pedestrians} (see \ref{app:relation}).
By using a rotational symmetry, the problem can be mapped to a random matrix model, which can
subsequently be solved by the standard matrix model methods \cite{randommat,realnum1,realnum2}. 
In fact the results of Section~\ref{sec:analytic} can be obtained from those in \cite{randommat}, 
as explained in \ref{app:relation}. 
However we expect that the field theoretical method employed in this paper may provide 
new insights and new applications in future analysis of the tensor eigenvalue/vector problems, which are still 
largely unexplored.
 }

This paper is organized as follows. In Section~\ref{sec:eveceig}, we define what we compute, 
the real eigenvalue/vector distributions for real symmetric order-three tensors with Gaussian
distributions. In Section~\ref{sec:bosferm}, we rewrite the problem as computation of 
a partition function of a zero-dimensional boson-fermion system with four-interactions. 
In Section~\ref{sec:fermion}, we determine the general form
which is obtained by integrating out the fermions in the partition function. In Section~\ref{sec:ai}, we perform 
the remaining bosonic integrals but one. In Section~\ref{sec:analytic}, we explicitly perform
the last bosonic integral for each case of $N\leq 8$, 
and obtain the explicit exact expressions of the distributions.
Interestingly, the apparently complicated bosonic integrals are feasible and the final results have 
simple expressions.    
In Section~\ref{sec:monte}, we perform some Monte Carlo simulations, and precise agreement with 
the exact expressions is obtained. 
In Section~\ref{sec:extrapolation}, we perform an extrapolation of the exact results for small $N$ to guess a
large-$N$ expression. We compare it with Monte Carlo simulations, and good agreement is obtained.
The last section is devoted to a summary and future prospects.

\section{Real eigenvector/value distributions of tensors}
\label{sec:eveceig}
In this paper, we restrict ourselves to the real symmetric order-three tensors, i.e.,
$C_{abc} \in \mathbb{R}\ (a,b,c=1,2,\ldots,N)$, satisfying $C_{abc}=C_{bac}=C_{bca}$,
as the simplest case.
As for eigenvalues/vectors of tensors,
there are a few similar but slightly different definitions in the literature \cite{Qi,lim,cart}.
In this paper we employ the definition that the real eigenvectors $v$ of a given $C$ are the 
non-zero solutions to
\[
C_{abc}v_b v_c=v_a,\ (v\neq 0,\ v\in \mathbb{R}^N),
\label{eq:evec}
\]
where the repeated indices are assumed to be summed over, as will be assumed in the rest of 
this paper. Then the distribution of the eigenvectors for a given $C$ is given by
\s[
\rho(v,C)&=\sum_{i=1}^{n_C} \delta^N (v-v^i) \\
&=| \det M | \prod_{a=1}^N \delta(v_a-C_{abc} v_b v_c) 
\label{eq:distevec}
\s]
for $v\neq 0$ under the volume measure $d^N v=\prod_{a=1}^N dv_a$, 
where $v^i\ (i=1,2,\ldots,n_C)$ are all the solutions to \eq{eq:evec}, 
$|\cdot |$ denotes the absolute value, and
\[
M_{ab}=\frac{\partial}{\partial v_a} \left( v_b -C_{bcd} v_c v_d \right)=\delta_{ab}-2 C_{abc}v_c.
\label{eq:defofm}
\]
Here the absolute value of the determinant, $|{\rm det} M|$, is the Jacobian associated to 
the change of the arguments of the $\delta$-functions performed in \eq{eq:distevec}.  

When the tensor $C$ has a Gaussian distribution, the real eigenvector distribution is
given by
\[
\rho(v)=A^{-1} \int_{\mathbb{R}^{\# C} } dC\, e^{-\alpha C^2} \rho(v,C), 
\label{eq:rhov}
\]
where $dC=\prod_{a\leq b \leq c=1}^N dC_{abc}$, $A=\int_{\mathbb{R}^{\# C}} dC \, e^{-\alpha C^2}$,
$C^2=C_{abc}C_{abc}$, $\alpha>0$,
and $\# C=N(N+1)(N+2)/6$,
which is the total number of the independent components of $C$.
   
It is worth commenting on the eigenvalue distribution corresponding to the real eigenvector
distribution above.
An eigenvalue $\zeta$ accompanied with a real eigenvector $v$ 
(a Z-eigenvalue in the terminology of \cite{Qi}) is defined by 
\[
C_{abc}w_b w_c = \zeta \, w_a\ (|w|=1,\ w\in \mathbb{R}^N),
\label{eq:zegeq}
\]
where $|w|=\sqrt{w_a w_a}$.
Comparing with \eq{eq:evec}, we obtain the relation, 
\[
\zeta=\frac{1}{|v|}.
\label{eq:eval}
\]
Therefore, by using \eq{eq:eval} and the fact that $\rho(v)$ is actually a function of $|v|$ because
of the rotational symmetry of the distribution, we obtain the real eigenvalue distribution,
\[
\rho_{\rm eig}(\zeta)=\rho(1/\zeta) S_{N-1} \zeta^{-N-1},
\label{eq:rhoz}
\]
where $1/\zeta$ in the argument abusively represents an arbitrary vector of size $1/\zeta$, and 
$S_{N-1}=2 \pi^{N/2}/\Gamma[N/2]$, the surface volume of a unit sphere in an $N$-dimensional space. 

{{\color{blue}} Here we would like to stress that we are specifically considering real eigenvalues/vectors only. 
Unlike real symmetric matrices, real symmetric tensors can also have complex eigenvalues/vectors 
by allowing complex solutions to the equation in \eq{eq:evec}. 
To include such complex solutions in our analysis, 
the expressions in this section need to appropriately be modified.
Note also that the connection to the $p$-spin spherical model explained in \ref{app:relation} is lost in this case. 
Therefore the end results for the complex case may have non-trivial differences from the real case.
}

\section{A zero-dimensional boson-fermion system}
\label{sec:bosferm}
In this section, we will rewrite \eq{eq:rhov} with \eq{eq:distevec} as a partition function
of a zero-dimensional boson-fermion system with four-interactions. 
An immediate obstacle in doing this is the presence of an absolute 
value in \eq{eq:distevec}, which is not an analytic function. To rewrite it in an analytic form, we take
\[
|\det M| =\lim_{\epsilon \rightarrow +0} \frac{\det (M^2+\epsilon I )}{\sqrt{\det (M^2+\epsilon I)}},
\label{eq:detm}
\]
where $I_{ab}=\delta_{ab}$ is an identity matrix, and the parameter $\epsilon$ is a positive small regularization 
parameter, which assures the convergence of the integrals below. 

{{\color{blue}}
Determinant factors like \eq{eq:detm} can be managed inside partition functions
of zero-dimensional field theoretical systems. 
This technique is common in supersymmetric approaches to disorder averaging in statistical physics
(see for instance \cite{fyodorov} and references therein). }
The numerator $\det (M^2+\epsilon I)$ can be rewritten as $\det(M^2+\epsilon I)=\int d\bpsi d\psi\, e^{\bpsi_a (M^2)_{ab}\psi_b+\epsilon \bpsi_a \psi_a}$ by 
introducing a fermion pair, $\bpsi$ and $\psi$ \cite{zinn}.
However, the exponent of this expression contains $C$ in a quadratic manner (see \eq{eq:defofm})
and is difficult to handle when we perform an integration over $C$ as will be done below.
Therefore, as was done previously in \cite{Sasakura:2022iqd}, 
we introduce another fermion pair, $\bvphi,\vphi$,
to rewrite the exponent in a form linear in $C$:
\[
\det ( M^2+\epsilon I)=(-1)^N \int d\bpsi d\psi d\bvphi d\vphi\, e^{-\bvphi \vphi-\bpsi M \vphi-\bvphi M \psi
+\epsilon \bpsi \psi },
\label{eq:m2}
\] 
where the contracted indices are suppressed for brevity: 
$\bvphi\vphi=\bvphi_a\vphi_a$, $\bpsi M \vphi=\bpsi_a M_{ab} \vphi_b$, and so on. 
The equality can be shown by noting that $\bvphi \vphi+\bpsi M \vphi+\bvphi M \psi=(\bvphi+\bpsi M)(\vphi+M \psi)
-\bpsi M^2 \psi$. 
In a similar manner we obtain
\[
\frac{1}{\sqrt{\det (M^2+\epsilon I)}}=\pi^{-N} \int_{\mathbb{R}^{2N}}
 d\phi d\sigma \, e^{-\sigma^2 -2 i \sigma M \phi -\epsilon \phi^2},
 \label{eq:sqrtm2}
\]
where $\sigma$ is a new bosonic variable, and $\sigma^2=\sigma_a \sigma_a$, and so on.

By using \eq{eq:detm}, \eq{eq:m2}, \eq{eq:sqrtm2} and a well-known formula, 
$\int_\mathbb{R} dx\, e^{i p x}=2\pi \delta(p)$, 
\eq{eq:rhov} with \eq{eq:distevec} can be rewritten as
\[
\rho(v)=\lim_{\epsilon\rightarrow +0}
A^{-1} (2 \pi)^{-N} \pi^{-N} (-1)^N \int dCd\lambda d\phi d\sigma d\bpsi d\psi d\bvphi d\vphi
\, e^{S_1},
\label{eq:rhov1}
\]
where 
\[
S_1=-\alpha C^2 +i \lambda_a (v_a-C_{abc}v_bv_c) 
-\sigma^2 -2 i \sigma M \phi -\epsilon \phi^2-\bvphi \vphi-\bpsi M \vphi-\bvphi M \psi
+\epsilon \bpsi \psi
\label{eq:s1}
\]
with a new bosonic variable $\lambda_a\ (a=1,2,\ldots,N)$ to rewrite the $\delta$-functions.
   
Next, let us perform the integrations over $C$ and $\lambda$ in the expression \eq{eq:rhov1}. Let us first
consider $C$. The terms containing $C$ in \eq{eq:s1} are
\[
S_C=-\alpha C^2 -i\, C_{abc}\lambda_a v_b v_c + 4 i C_{abc} v_a \sigma_b \phi_c + 2 C_{abc}v_a \bpsi_b \vphi_c
+2 C_{abc}v_a \bvphi_b \psi_c.
\]
Then we obtain
\[
\int_{\mathbb{R}^{\# C}} dC\, e^{S_C}= A\, e^{\delta S_C},
\]
where
\[
\delta S_C=\frac{1}{\alpha} \left(\frac{1}{6}
\sum_{s} \left( -\frac{i}{2} \lambda_{s_a} v_{s_b} v_{s_c} + 2 i v_{s_a}  \sigma_{s_b} \phi_{s_c}
+v_{s_a} \bpsi_{s_b} \vphi_{s_c} +v_{s_a} \bvphi_{s_b} \psi_{s_c}
\right)
\right)^2,
\label{eq:delsc1}
\]
where the summation is over all the permutations of $a,b,c$, which is necessary 
because $C$ is a symmetric tensor.
By explicitly expanding \eq{eq:delsc1}, we obtain
\s[
\delta S_C&=-\frac{|v|^4}{12 \alpha} \lambda_a B_{ab} \lambda_b -i \lambda_a (D_a+\tilde D_a)
+E_1+E_2+E_3,
\label{eq:delsc2}
\s]
where
\s[
B_{ab}&=\delta_{ab} +2 \hat v_a \hat v_b=I_{\perp\,ab}+3 I_{\parallel\,ab}, \\
D_a&=\frac{|v|^3}{3 \alpha} \left( \pbpsi \pvphi{}  +\pbvphi \ppsi{} \right)\hat v_a +
\frac{|v|^3}{3 \alpha} \left( 
\bpsi_a \pvphi{}+\pbpsi \vphi_{a}+\bvphi_a \ppsi{}+\pbvphi \psi_{a}
\right), \\
\tilde D_a&=\frac{2i}{\alpha} v_b v_c\cdot \frac{1}{6}
\sum_{s} v_{s_a} \sigma_{s_b} \phi_{s_c},
\\
E_1&=\frac{1}{\alpha}\left( 
\frac{1}{6} \sum_s \left( 
\bpsi_{s_a} \vphi_{s_b} v_{s_c}
+\bvphi_{s_a}\psi_{s_b} v_{s_c} 
\right)
\right)^2, \\
E_2&=
-\frac{4}{\alpha}
 \left( \frac{1}{6} \sum_s v_{s_a} \sigma_{s_b} \phi_{s_c} \right)^2, \\
E_3 &= \frac{4 i}{\alpha} 
(v_{a} \bpsi_{b} \vphi_{c} +v_{a} \bvphi_{b}\psi_{c} ) \cdot
\frac{1}{6} \sum_s v_{s_a} \sigma_{s_b} \phi_{s_c}.
\label{eq:delscterms}
\s]
Here $\hat v=v/|v|$, and $\pbpsi=\bar \psi_a \hat v_a$, etc., and 
$I_{\parallel}$ and $I_{\perp}$ are respectively the projection matrices to the parallel and transverse subspaces
to $\hat v$.

Next let us perform the integration over $\lambda$. Picking up the terms containing $\lambda$ (with no $C$) 
in \eq{eq:s1} and \eq{eq:delsc2}, we obtain
\[
S_{\lambda}=-\frac{|v|^4}{12 \alpha} \lambda_a B_{ab} \lambda_b +i \lambda_a (v_a-D_a-\tilde D_a).
\]
Considering that $B$ is a sum of the projection matrices as in \eq{eq:delscterms}, 
we obtain
\[
\int_{\mathbb{R}^N} d\lambda \, e^{S_\lambda}=|v|^{-2 N} (12 \pi \alpha)^\frac{N}{2} (\det B)^{-\frac{1}{2}} e^{\delta S_\lambda},
\]
where $\det B=3$ from \eq{eq:delscterms}, and 
\s[
\delta S_\lambda&=- 3 \alpha |v|^{-4} (v-D-\tilde D)_a B^{-1}_{ab} (v-D-\tilde D)_b \\
&=- \alpha |v|^{-2} +2 \alpha |v|^{-3} D_\parallel +2 \alpha |v|^{-3} \tilde D_\parallel \\
&\ \ \ \ -3 \alpha |v|^{-4} 
\left(D_\perp\cdot D_\perp +\frac{1}{3} D_\parallel^2+
2 D_\perp\cdot \tilde D_\perp + \tilde D_\perp^2+\frac{2}{3} D_\parallel \tilde D_\parallel+\frac{1}{3} \tilde D_\parallel^2
\right).
\label{eq:deltaslam}
\s]
Here we have used $B^{-1}=I_\perp+\frac{1}{3} I_\parallel$, and 
$D_\perp\cdot \tilde D_\perp =D_{\perp \, a} \tilde D_{\perp\, a}$, and so on,
where $X_\perp\ (X=D,\tilde D)$ denotes vector $X$ projected to the transverse subspace to $\hat v$.  
From \eq{eq:delscterms} the projections of $D$ and $\tilde D$ are more explicitly given by
\s[
D_\parallel &= \frac{|v|^3}{\alpha} \left(\pbpsi \pvphi+ \pbvphi \ppsi \right),\\
D_\perp&=\frac{|v|^3}{3 \alpha} \left( \tbpsi \pvphi+\pbpsi \tvphi{}+\tbvphi \ppsi{}+\pbvphi \tpsi \right), \\
\tilde D_\parallel &= \frac{2 i |v|^3}{\alpha} \sigma_\parallel \phi_\parallel,\\
\tilde D_\perp &= \frac{2 i |v|^3}{3 \alpha} \left( \sigma_\parallel \phi_\perp + \sigma_\perp \phi_\parallel\right). 
\s]
   
The first term in the second line of \eq{eq:deltaslam} is a constant contribution to the potential. The 
second and the third terms in the same line are corrections to the kinetic terms of the parallel components
of the fermions and the bosons, respectively. The other terms describe
 the four-interaction terms among the bosons and the fermions. Collecting 
the results above, we obtain the following intermediate expression,
\[
\rho(v)=\lim_{\epsilon\rightarrow +0}  3^{\frac{N-1}{2}} \pi^{-\frac{3 N}{2}} \alpha^\frac{N}{2} |v|^{-2 N} 
e^{-\frac{\alpha}{|v|^2} }(-1)^N \int d\phi d\sigma d\bpsi d\psi d\bvphi d\vphi \, e^{\tilde K_F+\tilde K_B 
+\tilde V_F+\tilde V_B+\tilde V_{BF}},
\]
where
\s[
\tilde K_F&=-\tbvphi \cdot \tvphi-\tbpsi \cdot \tvphi-\tbvphi\cdot \tpsi
+\epsilon \tbpsi \cdot \tpsi-\pbvphi \pvphi+\pbpsi \pvphi+\pbvphi \ppsi
+\epsilon \pbpsi \ppsi, \\
\tilde K_B&=-\tsig^2 -2 i \tsig\cdot \tphi -\epsilon \tphi^2-\psig^2 +2 i \psig \pphi -\epsilon \pphi^2, \\
\tilde V_F&=E_1-3 \alpha |v|^{-4} 
\left(D_\perp\cdot D_\perp +\frac{1}{3} D_\parallel^2\right), \\
\tilde V_B&=E_2-3 \alpha |v|^{-4} 
\left( \tilde D_\perp^2+\frac{1}{3} \tilde D_\parallel^2 \right), \\
\tilde V_{BF}&= E_3-3 \alpha |v|^{-4} 
\left(
2 D_\perp\cdot \tilde D_\perp +\frac{2}{3} D_\parallel \tilde D_\parallel 
\right).
\label{eq:intvs}
\s] 
Here $\tilde K_F,\tilde V_F$ contain only the fermions, $\tilde K_B,\tilde V_B$ only the bosons, and 
$\tilde V_{BF}$ is a mixture.
Note that the quadratic terms of the parallel components in $\tilde K_F$ and $\tilde K_B$ have been
corrected by the aforementioned terms
from the second line of \eq{eq:deltaslam}. 

Now we want to compute $\tilde V_B,\tilde V_F,\tilde V_{BF}$ more explicitly. 
The computation of $\tilde V_F$ in \eq{eq:intvs} is essentially the same as the derivation of
the four-fermi interaction terms in \cite{Sasakura:2022iqd} for $R=1$.
By noting that two of the terms in \cite{Sasakura:2022iqd} do not appear, because
$\tbpsi\cdot \tbpsi= \tbvphi\cdot \tbvphi=0$ (for $R=1$), we obtain
\s[
\tilde V_F&=-\frac{|v|^2}{6 \alpha} \left(
(\tbpsi\cdot \tvphi)^2+ (\tbvphi\cdot \tpsi)^2 
+2 \tbpsi\cdot \tbvphi \tvphi\cdot \tpsi+2 \tbpsi\cdot \tpsi{} \tbvphi\cdot \tvphi{} \right).
\label{eq:vf}
\s]
What seems surprising in this expression is that the parallel components of the fermions cancel out
from the interactions after all. 

As for $\tilde V_B$ and $\tilde V_{BF}$ in \eq{eq:intvs}, by using the results of the computations of $E_2$ and $E_3$ 
in \ref{app:e2e3}, we obtain
\s[
\tilde V_B&=-\frac{2 |v|^2}{3 \alpha} \left( \tsig^2 \tphi^2+(\tsig\cdot \tphi)^2 \right), \\
\tilde V_{BF}&=\frac{2 i |v|^2}{3 \alpha} \left( 
\tbpsi\cdot\tsig\, \tvphi\cdot \tphi+\tbvphi\cdot \tsig\,\tpsi\cdot\tphi+
\tbpsi\cdot \tphi\,\tvphi\cdot \tsig+ \tbvphi\cdot \tphi\, \tpsi\cdot\tsig
\right).
\label{eq:vbbf}
\s]
We again find the surprising fact that no parallel components appear in $\tilde V_B$ and $\tilde V_{BF}$.

Because the parallel components only exist in $\tilde K_B,\tilde K_F$ and do not interact, 
these can trivially be integrated out,
that generates the overall factors of $\pi$ and $-1$ for the bosons and fermions, respectively.
Therefore we finally obtain
\[
 \rho(v)=\lim_{\epsilon\rightarrow +0}  3^{\frac{N-1}{2}} \pi^{-\frac{3 N}{2}+1} \alpha^\frac{N}{2} |v|^{-2 N} 
e^{-\frac{\alpha}{|v|^2} }(-1)^{N-1} \int d\tphi d\tsig d\tbpsi d\tpsi d\tbvphi d\tvphi \, e^{\tilde K^\perp_F
+\tilde K^\perp_B +\tilde V_F+\tilde V_B+\tilde V_{BF}},
\label{eq:finalrho}
\]
where 
\s[
\tilde K^\perp_F&=-\tbvphi \cdot \tvphi-\tbpsi \cdot \tvphi-\tbvphi\cdot \tpsi
+\epsilon \tbpsi \cdot \tpsi, \\
\tilde K^\perp_B&=-\tsig^2 -2 i \tsig\cdot \tphi -\epsilon \tphi^2, 
\s]
and $\tilde V_F,\tilde V_B,\tilde V_{BF}$ are given in \eq{eq:vf} and \eq{eq:vbbf}.
   
\section{Integrations over fermions} 
\label{sec:fermion}
In the expression \eq{eq:finalrho} the variables projected to the transverse directions, i.e., $\bpsi_\perp$, etc., 
just represent $N-1$-dimensional degrees of freedom with no other restrictions. Therefore we can simply
regard them as $N-1$-dimensional variables from the beginning.
The external variable, the vector $v$, only appears in the coupling constants of 
the four-interactions through $|v|$.
Therefore we can simply write 
\[
\rho(v)=\lim_{\epsilon\rightarrow +0}  3^{\frac{N-1}{2}} \pi^{-\frac{3 N}{2}+1} \alpha^\frac{N}{2} |v|^{-2 N} 
e^{-\frac{\alpha}{|v|^2} }(-1)^{N-1} \int_{N^*} d\phi d\sigma d\bpsi d\psi d\bvphi d\vphi \, e^{K_F
+K_B +V_F+V_B+V_{BF}},
\label{eq:newrho}
\] 
where all the variables are $N-1$-dimensional, and 
\s[
K_F&=-\bvphi \cdot \vphi-\bpsi \cdot \vphi-\bvphi\cdot \psi
+\epsilon \bpsi \cdot \psi, \\
K_B&=-\sigma^2 -2 i \sigma\cdot \phi -\epsilon \phi^2, \\
V_F&=-\frac{|v|^2}{6 \alpha} \left(
(\bpsi\cdot \vphi)^2+ (\bvphi\cdot \psi)^2 
+2 \bpsi\cdot \bvphi\, \vphi\cdot \psi+2 \bpsi\cdot \psi \, \bvphi\cdot \vphi{} \right),\\
 V_B&=-\frac{2 |v|^2}{3 \alpha} \left( \sigma^2 \phi^2+(\sigma\cdot \phi)^2 \right), \\
 V_{BF}&=\frac{2 i |v|^2}{3 \alpha} \left( 
\bpsi\cdot\sigma\, \vphi\cdot \phi+\bvphi\cdot \sigma\,\psi\cdot\phi+
\bpsi\cdot \phi\,\vphi\cdot \sigma+ \bvphi\cdot \phi\, \psi\cdot\sigma
\right).
\label{eq:newkv}
\s]
Note that, for simplicity, we are abusively using the same notations of the variables as those in Section~\ref{sec:bosferm}
with a different dimension, and, to indicate this difference, 
the symbol $N^*$ is attached to the integral symbol in \eq{eq:newrho}.

To compute $\rho(v)$ explicitly, we first perform the fermionic integrations.
The integrand in \eq{eq:newrho} can be expanded in the fermions. 
A useful property in this expansion is that the expansion in $V_{BF}$ stops at the fourth order:
\[
e^{V_{BF}}=\sum_{n=0}^4 \frac{1}{n!} \left(V_{BF}\right)^n.
\label{eq:expbf}
\]
This can be proven as follows. The fermions in $V_{BF}$ are projected to $\phi$ or $\sigma$.
Therefore $V_{BF}$ contains only eight independent fermions in total, i.e., $\bpsi\cdot \phi,\bpsi\cdot \sigma,\psi\cdot \phi,\psi\cdot \sigma,\bvphi\cdot \phi,
\bvphi\cdot \sigma,\vphi\cdot \phi,\vphi\cdot \sigma$. Since each term of $V_{BF}$ contains 
two of these fermions and products of more than eight of these fermions vanish, we obtain
\[
(V_{BF})^n=0\hbox{ for }n>4.
\label{eq:vbf4}
\]

We also notice that, when $\sigma=\pm \phi$,  only four of these fermions are independent. Therefore,
\[
(V_{BF})^n=0 \hbox{ for } n>2,\hbox{ when }\sigma=\pm \phi.
\label{eq:vbf2}
\]

Now we want to determine the functional forms of the fermionic integrals of each summand in \eq{eq:expbf}:
\[
\beta_n=\frac{(-1)^{N-1}}{n!} \int_{N^*} d\bpsi d\psi d\bvphi d\vphi\, (V_{BF})^n e^{K_F +V_F}\ \ \ (n\leq 4).
\]
From the form of $V_{BF}$ and the $O(N-1)$ symmetry, $\beta_n$ should be
a polynomial function of $\sigma^2, \phi^2, \sigma\cdot \phi$, and its order in $\sigma,\phi$ should be $2n$.
In addition, each term of the polynomial function should 
contain equal numbers of $\sigma$ and $\phi$, and 
the polynomial function should also be invariant under the interchange $\sigma\leftrightarrow \phi$
as a whole.

From the above considerations,  we uniquely obtain for $n=1$
 \[
\beta_1=a_1 \sigma\cdot \phi.
\] 
Here $a_1$ is a function of $\epsilon$ and $|v|^2/\alpha$, and all $a_i$ below are also so. 

For $n=2$, we have two possibilities,
\[
\beta_2=a_2 (\sigma\cdot \phi)^2+a_3 \sigma^2 \phi^2.
\]

As for $n=3$, $\beta_3$ must vanish for $\sigma=\pm \phi$ due to \eq{eq:vbf2}. Considering also the 
other conditions mentioned above, we uniquely obtain
\[
\beta_3&=a_4 \left( \sigma^2 \phi^2 \sigma\cdot \phi- (\sigma\cdot \phi)^3 \right).
\] 
   
As for $n=4$, we need not only \eq{eq:vbf2} but also the following property:
Taking the derivatives of $(V_{BF})^4$ with respect to $\sigma,\phi$ three times or less must vanish, when 
$\sigma=\pm \phi$. This can be proven as follows.
Obviously, $(V_{BF})^4$ is proportional to the product of the eight projected fermions, 
$\bpsi\cdot \phi,\bpsi\cdot \sigma,\ldots$. 
After taking the derivatives of this three times or less with respect to
 $\sigma,\phi$, each term contains at least five of the eight projected fermions. 
Therefore all these terms vanish for $\sigma=\pm \phi$ because of the same reason for \eq{eq:vbf2}.
Now using this property, we can uniquely determine
\[
\beta_4=a_5 \left( \sigma^2 \phi^2-(\sigma\cdot \phi)^2 \right)^2.
\]
 
Collecting all the results above, we conclude that the fermionic integrations of the partition
function has the following general form,
\s[
&\int_{N^*} d\bpsi d\psi d\bvphi d\vphi \, e^{K_F+V_F+V_{BF}}\\
&=
a_0+a_1 \sigma\cdot \phi+a_2 (\sigma\cdot \phi)^2+a_3 \sigma^2 \phi^2 
+a_4 \left( \sigma^2 \phi^2 \sigma\cdot \phi- (\sigma\cdot \phi)^3 \right)
+a_5 \left( \sigma^2 \phi^2-(\sigma\cdot \phi)^2 \right)^2.
\label{eq:fermiai}
\s]
Here $a_i$ are generally functions of $\epsilon$ and $|v|^2/\alpha$, 
but $\epsilon$ in them turn out to simply disappear in the $\epsilon \rightarrow +0$ limit 
in \eq{eq:newrho} without any essential roles.
Therefore in the following sections we ignore it by just putting $\epsilon=0$ in $a_i$. 

\section{Integrations over bosons}
\label{sec:ai}
Assuming the form \eq{eq:fermiai} of the fermionic integrations, we perform the bosonic integrations
but one in this section. 

Let us first generalize the bosonic kinetic term by introducing new parameters $\epsilon_i\ (i=1,2,3)$:
\[
K^{\epsilon}_{B}=-\epsilon_1 \sigma^2 -2 i \epsilon_2\, \sigma\cdot \phi -\epsilon_3 \phi^2,
\]
where $\epsilon_1,\epsilon_3>0$ are assumed for the convergence of the bosonic integration.
The original kinetic term in \eq{eq:newkv} corresponds to 
$\epsilon_1=\epsilon_2=1,\ \epsilon_3=\epsilon$.

Using this new kinetic term, \eq{eq:newrho}, and \eq{eq:fermiai}, $\rho(v)$ can be expressed as 
\s[
&\rho(v)= 3^{\frac{N-1}{2}} \pi^{-\frac{3 N}{2}+1} \alpha^\frac{N}{2} |v|^{-2 N} e^{-\frac{\alpha}{|v|^2} }\, G_N,
\label{eq:rhovbyh}
\s]
where  
\s[ 
G_N=\left( a_0+a_1 D_2 +a_2 D_2^2 +a_3 D_1+a_4 (D_1 D_2 -D_2^3)+a_5 (D_1-D_2^2)^2\right)
\left.
\int_{N^*} d\sigma d\phi\, e^{K^{\epsilon}_B+V_B} \right|_{\epsilon_1=\epsilon_2=1 \atop \epsilon_3=+0},
\label{eq:rhovdel}
\s] 
with $D_1,D_2$ being the following partial derivative operators, 
\s[
D_1&=\frac{\partial^2}{\partial \epsilon_1 \partial \epsilon_3}, \\
D_2&=-\frac{1}{2 i} \frac{\partial}{\partial \epsilon_2}.
\s]
   
The bosonic integration in \eq{eq:rhovdel} does not seem to be fully integrable, but it has a simpler expression.
Since $\sigma$ appears at most 
quadratically in $K_B^\epsilon+V_B$, the $\sigma$ integration can be performed:
\[
\int_{N^*} d\sigma\,  e^{K^\epsilon_B+V_B}=\pi^{\frac{N-1}{2}} (\epsilon_1+8 z \phi^2)^{-\frac{1}{2}} (\epsilon_1+4 z \phi^2)^{\frac{N-2}{2}} \exp \left( -\frac{(\epsilon_2^2+\epsilon_3 (\epsilon_1+8 z \phi^2))\phi^2} {\epsilon_1+8 z \phi^2}\right),
\label{eq:intsig}
\]
where for brevity we have introduced 
\s[
z=\frac{|v|^2}{6 \alpha}.
\label{eq:defz}
\s]
\eq{eq:intsig} does not depend on the angular directions of $\phi$, and therefore the integration over these
produces the spherical volume, $2 \pi^{(N-1)/2} |\phi|^{N-2}/\Gamma[(N-1)/2]$. 
Then, after applying the derivative operators in \eq{eq:rhovdel} and performing a replacement of 
variable, $|\phi |=\sqrt{x}/\sqrt{1-8 z x}$, we obtain
\s[
G_N=\frac{\pi^{N-1}}{4 \, \Gamma\left[\frac{N-1}{2}\right]}
 \int_0^\frac{1}{8 z}& dx  \, e^{-x}x^{\frac{N-3}{2}} (1-4 z x)^{-\frac{N+2}{2}} \\
&\cdot (4 a_0 + 2 (-2 i a_1 + a_2 +  (N-1) a_3 ) x\\
&\ \ \  + (8 i z a_1- (3 + 4 z)(a_2+a_3)  - i (N-2) a_4   )x^2+ 8 z  (a_2  + a_3 ) x^3 ),
\label{eq:defg}
\s]
where we have used 
\[
 64 z^2 a_0  - 8 i  z a_1+ 
(4 z-1)a_2  +  (-1 - 4 z+ 8 (N-1) z)a_3- i (N-2) a_4 + N (N-2) a_5=0
\label{eq:relationai}
\]
to delete $a_5$. As we will see in Section~\ref{sec:analytic}, 
\eq{eq:relationai} holds for all the cases we consider (namely, $N\leq 8$). 
In fact this relation is essentially important, because otherwise 
the integrand has an extra factor $1/(1 - 8 z x)$, and the feasibilty of the integration over $x$ which 
will be performed in Section~\ref{sec:analytic} would become unclear.

\section{Analytic expressions}
\label{sec:analytic}
In this section, we explicitly compute the integration \eq{eq:defg} for each case of $N \leq 8$ to obtain
exact analytic expressions. What is surprising is that the seemingly difficult integrations
can be done explicitly.
The apparent reason is that the integrand of \eq{eq:defg} 
turns out to be a sum of a total derivative and a simple integrable function. 

\subsection{$N=1$}
\label{subsec:neq1}
This case is trivial, since the integration in \eq{eq:newrho} can just be ignored. We obtain
\[
\rho_{N=1}(v)=\pi^{-\frac{1}{2}} \alpha^{\frac{1}{2}} |v|^{-2} e^{-\frac{\alpha}{|v|^2}}.
\]
Since $v=1/C$ for $N=1$ from \eq{eq:evec}, $\rho_{N=1}(v)\, dv$ is indeed equivalent to
$\alpha^{1/2} \pi^{-1/2} e^{-\alpha C^2} dC$, which is the Gaussian distribution of $C$.

\subsection{$N=2$}  
In this case, since the variables in \eq{eq:newrho} are all one-dimensional, 
we can ignore the indices. Then we immediately obtain
\s[
V^{N=2}_F&=-4 z \, \bpsi \psi \bvphi \vphi,\\
 V^{N=2}_B&=-8 z\, \sigma^2 \phi^2, \\
 V^{N=2}_{BF}&=8 i z\, \sigma \phi  \left( 
\bpsi \vphi+\bvphi \psi \right).
\s]
Hence, by explicitly expanding the integrand, we obtain,
\[
\int_{N^*} d\bpsi d\psi d\bvphi d\vphi\,  e^{K_F^{N=2} +V_F^{N=2}+V_{BF}^{N=2}}
=1+4 z -16 i z \sigma \phi -64 z^2 \sigma^2 \phi^2.
\]
This determines 
\s[
&a_0=1+4 z , \\
&a_1=-16 i z, \\
&a_2= -64 z^2, \\
&\hbox{Others}=0,
\label{eq:neq1ai}
\s]
by comparing with \eq{eq:fermiai}.
This indeed satisfies \eq{eq:relationai}. Putting \eq{eq:neq1ai} into \eq{eq:defg}, we obtain
\s[
G_{N=2}&=\sqrt{\pi}
 \int_0^\frac{1}{8 z} dx  \, e^{-x}x^{-\frac{1}{2}} (1+4 z -8zx) \\
 &=\sqrt{\pi}\left( 
 (1+4 z) \gamma\left[ \frac{1}{2},\frac{1}{8z} \right]-8 z \gamma \left[ \frac{3}{2},\frac{1}{8z}\right]
 \right) \\
 &= \sqrt{\pi} \left(
 \gamma \left[ \frac{1}{2} , \frac{1}{8z} \right]+\sqrt{8z}\, e^{-\frac{1}{8z}}
 \right),
 \label{eq:hneq2}
\s] 
where the lower incomplete gamma function $\gamma[\cdot,\cdot]$ is defined by
\[
\gamma \left[a,y\right]=\int_0^y dt\, t^{a-1}\, e^{-t},
\]
and we have used its property,
\[
\gamma\left[ a+1,y \right] =a \gamma\left[a,y\right] -y^a e^{-y}.
\]
The lower incomplete gamma function with index $1/2$ in \eq{eq:hneq2} is related to the error function by
\[
\gamma\left[\frac{1}{2},y\right]=\sqrt{\pi}\, \hbox{erf}\left[\sqrt{y}\right].
\]
Therefore, from \eq{eq:defz} and \eq{eq:hneq2}, 
$G_{N=2}$ is represented by a combination of polynomial, exponential, and error functions of $|v|$.
This is common to the other cases of $N$ shown below.   
   
\subsection{$N=3$}
\label{subsec:neq3}
The fermionic integration of \eq{eq:fermiai} seems too complicated to perform by hand. 
Rather, we use a Mathematica package for Grassmann variables \cite{grassmann}.
The result is 
\s[
a_0&=1 + 4 z + 28 z^2, \\
a_1&=  -16 i (z + 2 z^2), \\
a_2&=  -32 (3 z^2 + 2 z^3), \\
a_3&=  -32 (-z^2 + 6 z^3),  \\
a_4 &= -256 i z^3, \\
a_5  &= 256 z^4.
\s] 
This indeed satisfies \eq{eq:relationai}. Then, by putting this into \eq{eq:defg}, we obtain
\s[
G_{N=3}=\pi^2 \int_0^\frac{1}{8z} dx\, \frac{ e^{-x}}{(1-4 z x)^\frac{5}{2}}
\big(&
1+4z+28 z^2-16 z  (1 + 3 z + 14 z^2)x \\
&+ 16 z^2 (5 + 16 z + 16 z^2)x^2 - 128 z^3 (1 + 4 z) x^3
\big).
\label{eq:intg3}
\s]
A surprising fact is that the integrand in \eq{eq:intg3} is actually a total derivative, and we therefore obtain
\s[
G_{N=3}&=-\pi^2 \int_0^\frac{1}{8z} dx\, \frac{d}{dx} \left( \frac{e^{-x}}{(1-4zx)^\frac{3}{2}} \left(1-2z-4z(3+4z) x +32 z^2 (1+4z) x^2 \right)
\right) \\
&=\pi^2 \left( 1-2z +4 \sqrt{2} z e^{-\frac{1}{8z}}\right).
\s]
   
\subsection{Larger odd $N$}    
\label{subsec:odd}
The strategy taken in Section~\ref{subsec:neq3} can be generalized for larger odd $N$ 
in the following manner. We first compute the $a_i\, (i=0,1,\ldots,5)$, which are listed for $N=5,7$ 
in \ref{app:odd}, by using the aforementioned Mathematica package. They indeed satisfy \eq{eq:relationai}.
Then, by putting them into \eq{eq:defg}, we obtain $G_N$. 
Similarly to the case of $N=3$ in \ref{subsec:neq3}, what we find is that the integrand of $G_N$ is 
 a total derivative of the following form for $N=5,7$:
\[
G_{N:{\rm odd}}=\frac{\pi^{N-1}}{4 \, \Gamma\left[\frac{N-1}{2}\right]}
 \int_0^\frac{1}{8 z}& dx  \, \frac{d}{dx} \frac{e^{-x} \sum_{n=0}^\frac{N+1}{2} b_n x^n}{(1-4 z x)^\frac{N}{2}},
\]   
where $b_n\, (n=0,1,\ldots, (N+1)/2)$ are some polynomial functions of $z$.  The explicit
forms of $b_n$ are given in \ref{app:odd}.
Therefore, we obtain
\[
G_{N:{\rm odd}}=\frac{\pi^{N-1}}{4 \, \Gamma\left[\frac{N-1}{2}\right]}
\left( 2^\frac{N}{2}  e^{-\frac{1}{8 z}} \sum_{n=0}^\frac{N+1}{2} \frac{b_n}{(8 z)^n}-b_0\right).
\]

By using the $b_i$ in \ref{app:odd}, the explicit expressions of $G_N$ for $N=5,7$ are given by
\s[
G_{N=5}&=\pi^4 \left(1 - 12 z + 12 z^2 + \sqrt{2} e^{-\frac{1}{8z}} (1 + 12 z + 12 z^2)\right),
\\
G_{N=7}&=\pi^6 \left(1 - 30 z + 180 z^2 - 120 z^3 + 
   \frac{ \sqrt{2} e^{-\frac{1}{8z}}}{8 z} (1 + 8 z + 120 z^2 - 480 z^3 + 2640 z^4)\right).
\s]

Let us lastly comment on our equipment. 
The computations were done on a machine which had a Xeon W2295 (3.0GHz, 18 cores), 
128GB DDR4 memory, and Ubuntu 20 as OS. 
The computation of $a_i$ quickly takes longer time as $N$ becomes larger. $G_{N=8}$, which appears
in the next subsection, was the largest feasible case, while we failed to obtain
$G_{N=9}$ seemingly because of a memory shortage.

\subsection{Larger even $N$}    
The difference from the odd case of Section~\ref{subsec:odd} is that the integrand of $G_N$ is 
a sum of a total derivative and a simple integrable term:
\[
G_N=\frac{\pi^{N-1}}{4 \, \Gamma\left[\frac{N-1}{2}\right]}
 \int_0^\frac{1}{8 z}& dx \, \left(
 c_0 \,  x^{-\frac{1}{2}} e^{-x}
 +\frac{d}{dx} \frac{x^\frac{1}{2} e^{-x} \sum_{n=0}^\frac{N}{2} b_n x^n}{(1-4 z x)^\frac{N}{2}}
 \right).
\]  
Then, by doing the integration, we obtain
\[
G_N=\frac{\pi^{N-1}}{4 \, \Gamma\left[\frac{N-1}{2}\right]} \left(
c_0 \gamma\left[\frac{1}{2},\frac{1}{8z} \right] +2^\frac{N}{2} e^{-\frac{1}{8z}} \sum_{n=0}^\frac{N}{2}  \frac{b_n}{(8z)^{n+\frac{1}{2}}}
\right).
\]
The lists of $a_i,b_i,c_0$ for $N=4,6,8$ are given in \ref{app:even}. $a_i$ indeed satisfy \eq{eq:relationai}.
By putting the values of $b_i,c_0$ we obtain the explicit forms of $G_N$ as
\s[
G_{N=4}&=\pi^{\frac{5}{2}} \left(6 \sqrt{2} e^{-\frac{1}{8z}} \sqrt{z} (1 + 2 z) + (1 - 6 z) 
\,\gamma\left[\frac{1}{2},\frac{1}{8 z}\right]\right), \\
G_{N=6}&=
 \pi^\frac{9}{2} \left(\frac{2 \sqrt{2} e^{-\frac{1}{8z}} (1 + 15 z + 180 z^3)}{3\sqrt{z}} + 
    (1 - 20 z + 60 z^2)\, \gamma\left[\frac{1}{2},  \frac{1}{8 z}\right]\right),\\
G_{N=8}&= \pi^{\frac{13}{2}} \Bigg( \frac{
   \sqrt{2}
     e^{-\frac{1}{8z}} (1 + 210 z^2 - 2100 z^3 + 12600 z^4 + 25200 z^5)}{
   15 z^\frac{3}{2}} \\
   &\hspace{4cm}+
    (1 - 42 z + 420 z^2 - 840 z^3) \,\gamma\left[\frac{1}{2},\frac{1}{8 z}\right]
    \Bigg).
    \label{eq:hforeven}
\s]

\section{Comparison with Monte Carlo simulations}
\label{sec:monte}
In this section, we compare the results in Section~\ref{sec:analytic} with Monte Carlo simulations.
The procedure is basically the same as that used in \cite{Sasakura:2022zwc,Sasakura:2022iqd}. 
To make this paper self-contained, however, we review the method below.

The eigenvector equation \eq{eq:evec} is a system of polynomial equations and it can be solved 
by an appropriate polynomial equation solver, unless $N$ is too large. 
We use Mathematica 13 for this purpose.
It gives generally complex solutions to the equation \eq{eq:evec}, and we pick up only real ones, 
since we are only counting real eigenvectors (or Z-eigenvalues). Whether this method covers all the real solutions 
or not can be checked by whether the number of generally complex solutions obtained for each $C$
by a polynomial equation solver agrees with the known number $2^N-1$ 
\cite{cart}\footnote{In fact, for large $N$, Mathematica 13 seems to miss a few solutions for some $C$. 
We have not pursued the reason for that, but the missing portion is $\sim 10^{-4}$ even for our 
largest case of $N=16$, and is statistically irrelevant in the present study.}.    

With the above method of solving the eigenvector equation \eq{eq:evec}, 
our procedure of Monte Carlo simulation is given as follows.
\begin{itemize}
\item
Randomly generate a real symmetric tensor $C$ with components, 
$C_{ijk}=\sigma/\sqrt{d(i,j,k)} \ (1\leq i \leq j\leq k\leq N)$, where $\sigma$ has a normal distribution with mean value zero 
and standard deviation one, and $d(i,j,k)$ is a degeneracy factor, 
\s[
d(i,j,k)=\left\{ 
\begin{array}{ll}
1, & i=j=k \\
3 ,& i=j\neq k ,\,i\neq j=k,\, k=i\neq j \\
6 ,& i\neq k \neq j \neq i
\end{array}
\right.
.
\s]
This random generation of $C$ corresponds to $\alpha=1/2$ in \eq{eq:rhov},
since
\s[
C^2=C_{abc}C_{abc}=\sum_{i\leq j \leq k=1}^N d(i,j,k) \, C_{ijk} ^2
\s]
due to $C$ being a symmetric tensor.
\item
Compute the real eigenvectors of a generated $C$ by the aforementioned method.
\item
Store each size $|v|$ of all the real eigenvectors. 
\item
Repeat the above processes.
\end{itemize}
By this procedure, we obtain a sequence of sizes of real eigenvectors, $|v|_i\ (i=1,2,\ldots,L)$.
Then the size distribution of real eigenvectors can be obtained by
\[
\rho^{\rm MC}_{\rm size} ((k+1/2)\delta v)=\frac{1}{ \delta v N_C}\sum_{i=1}^L  
\theta(k \delta v <|v|_i\leq (k+1)\delta v),
\label{eq:rhosim}
\]
where $\delta v$ is a bin size, $N_C$ is the total number of randomly generated $C$, $k=0,1,2,\ldots$,
and $\theta(\cdot)$ is a support function which takes 1 if the inequality of the argument
is satisfied, but zero otherwise. For the comparison with the analytical results obtained 
in Section~\ref{sec:analytic}, \eq{eq:rhosim} should be compared with the size distribution 
\[
\rho_{\rm size}(|v|)= \rho(v) S_{N-1} |v|^{N-1}, 
\label{eq:rhosize}
\]
where $S_{N-1}=2 \pi^{N/2}/\Gamma[N/2]$ 
denotes the surface volume of a unit sphere in an $N$-dimensional space, 
and the vector $v$ in the argument of $\rho(\cdot)$ 
is an arbitrary vector of size $|v|$ due to the rotational symmetry.

Since a Z-eigenvalue is related with the size of an eigenvector by the relation \eq{eq:eval}, 
the real (or Z-) eigenvalue distribution is given by
\[
\rho^{\rm MC}_{\rm eig} ((k+1/2)\delta \zeta)=\frac{1}{ \delta \zeta N_C}\sum_{i=1}^L  
\theta(k \delta \zeta <1/|v|_i\leq (k+1)\delta \zeta),
\label{eq:mcz}
\]
where $\delta \zeta$ is a bin size.
This quantity should be compared with \eq{eq:rhoz}.

\begin{figure}
\begin{center}
\hfil
\includegraphics[width=7cm]{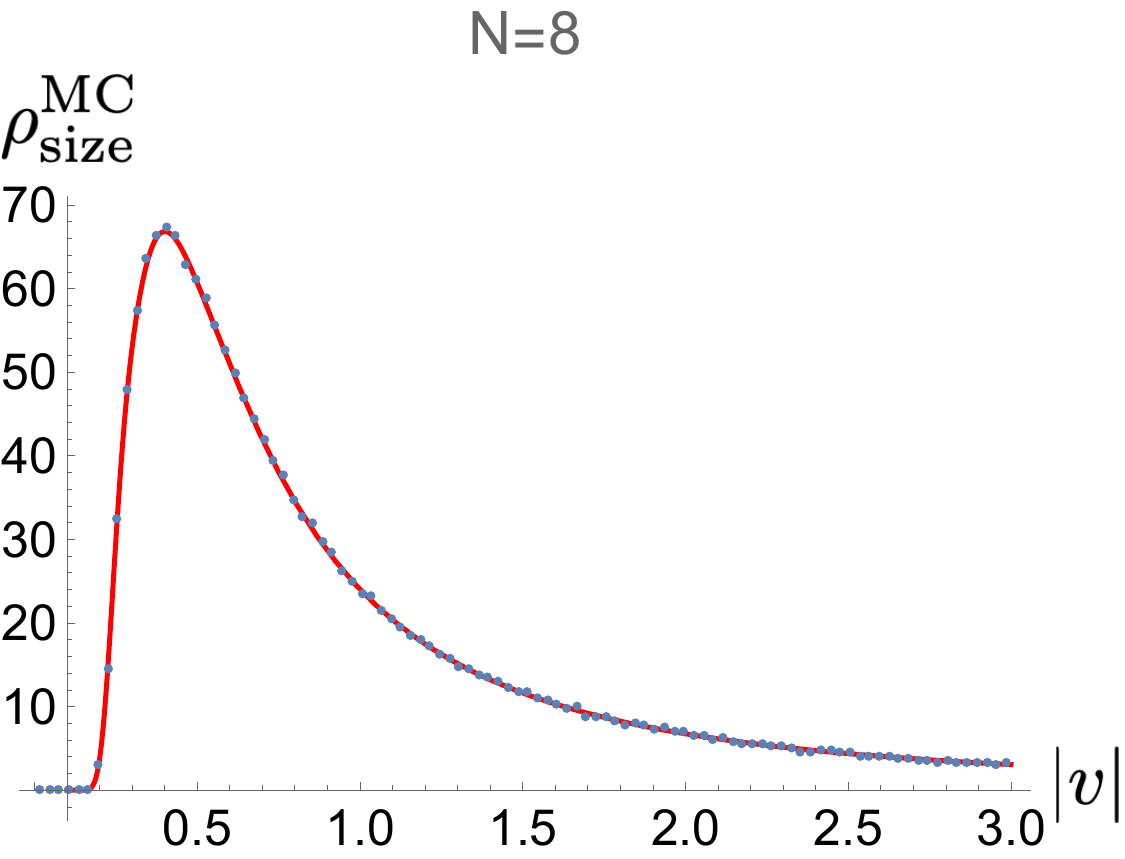}
\hfil
\includegraphics[width=7cm]{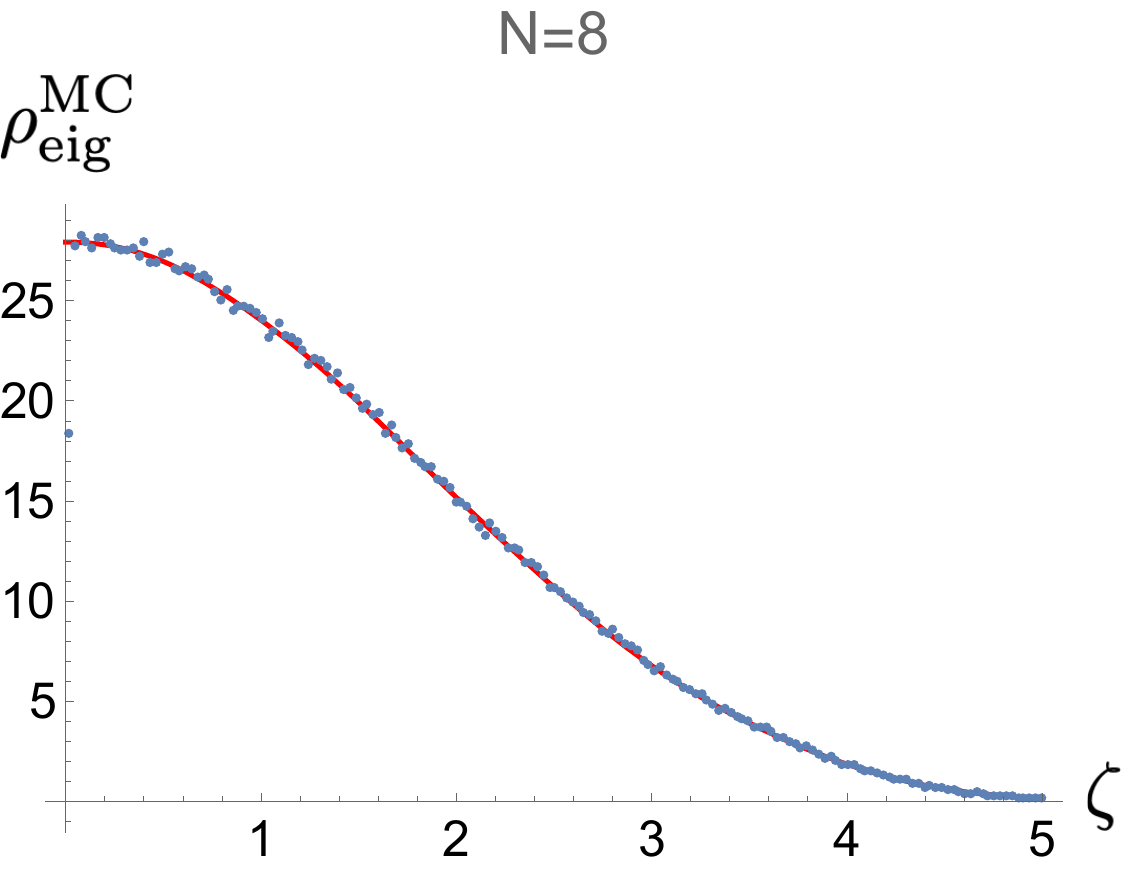}
\hfil
\caption{The results of the Monte Carlo simulation for $N=8, N_C=10000$
are compared with the analytic expressions. 
Left: The eigenvector size distribution. The Monte Carlo result \eq{eq:rhosim} with $\delta v=0.03$ (dots) is compared with
\eq{eq:rhosize} using \eq{eq:rhovbyh} and $G_{N=8}$ in \eq{eq:hforeven} 
(solid line).
Right: The eigenvalue distribution. The Monte Carlo result \eq{eq:mcz} with $\delta \zeta=0.03$ (dots) is compared with 
the analytic expression through \eq{eq:rhoz}.
\label{fig:mc}
}
\end{center}
\end{figure}

In Figure~\ref{fig:mc}, the Monte Carlo results are compared with the analytic expressions for $N=8$.
They agree precisely. Similar precise agreement has been obtained for all $N\leq 8$.

\section{Extrapolation to general $N$}
\label{sec:extrapolation}
In this section, we point out a few patterns which exist in the expressions of the distributions 
derived for small $N$ in Section~\ref{sec:analytic}, 
and guess an extrapolation to general $N$. 
A motivation for doing this is to improve the large-$N$ expression previously obtained by an approximation
using a Schwinger-Dyson equation \cite{Sasakura:2022iqd}. The issue of the previous result was that, 
while the functional form agreed well with the numerical simulation for large-$N$, the overall factor did not. 
In this section, using the extrapolation, we will guess the overall factor for general $N$ and find good agreement 
with Monte Carlo simulations.

After a thought one notices that $G_N$ for even $N$ (namely, $N=2,4,6,8$) in Section~\ref{sec:analytic} 
can be expressed by the following general form:
\s[
G_{N:{\rm even}}=&\pi^{N-\frac{3}{2}} z^\frac{N-1}{2} H_{N-1} \left[ \frac{1}{2 \sqrt{z}}\right] \gamma\left[ \frac{1}{2},\frac{1}{8z} \right]\\
&+\sqrt{2} \pi^{N-\frac{3}{2}} \frac{N!}{\frac{N}{2}!} z^{\frac{N-1}{2}} e^{-\frac{1}{8z}}
\left(1+\frac{d_1}{z}+\frac{d_2}{z^2}+\cdots+\frac{d_{N-3}}{z^{N-3}}\right),
\label{eq:generalG}
\s]
where $H_n[\cdot]$ are Hermite polynomials, $d_i$ are some coefficients generally depending on $N$, 
and specifically 
\[
d_1=\frac{1+(-1)^\frac{N}{2}}{4}.
\label{eq:d1}
\]
As for $d_i\ (i\geq 2)$, we could not find reasonably simple functions of $N$.

Let us discuss the real eigenvalue distribution for large-$N$, assuming \eq{eq:generalG} with \eq{eq:d1}.  
Because of the relations \eq{eq:eval} and \eq{eq:defz}, and the fact that the major part of the distribution is around $\zeta\sim0$
as in Figure~\ref{fig:mc}, 
we are interested in a $1/z$ expansion of \eq{eq:generalG} with \eq{eq:d1}.
By explicitly doing this one obtains
\[
G_N\sim \sqrt{2} \pi^{N-\frac{3}{2}} \frac{\Gamma[N+1]}{\Gamma\left[\frac{N}{2}+1\right]} z^{\frac{N-1}{2}} \left( 1+\frac{1}{8z} + \cdots \right),
\label{eq:gnasymp}
\]
where the factorials are replaced by Gamma functions to also be applicable for odd $N$ below.
By combining with the previous result in \cite{Sasakura:2022iqd}
that the large-$N$ eigenvalue distribution is given by a Gaussian function of $|v|$, 
one could assume that the expansion in $1/z$ of \eq{eq:gnasymp} 
comes from the expression, 
\[
G_N\sim \sqrt{2} \pi^{N-\frac{3}{2}} \frac{\Gamma[N+1]}{\Gamma\left[\frac{N}{2}+1\right]} z^{\frac{N-1}{2}} e^{\frac{1}{8z}}.
\]
Putting this into \eq{eq:rhoz} using \eq{eq:eval}, \eq{eq:rhovbyh} and \eq{eq:defz}, we obtain
\[
\rho_{\rm eig} (\zeta)\sim 2^{-\frac{N}{2}+2} \alpha^\frac{1}{2} \pi^{-\frac{1}{2}} \frac{\Gamma[N+1]}{\Gamma\left[\frac{N}{2}+1\right] \Gamma\left[\frac{N}{2}\right]} e^{-\frac{\alpha}{4}\zeta^2},
\label{eq:app}
\]
which indeed is a Gaussian distribution. While the coefficient $\alpha/4$ in the exponent 
indeed agrees with the previous result in \cite{Sasakura:2022iqd}, the overall factor is different. 
{{\color{blue}} By integrating over $\zeta$, one can obtain the mean total number of real eigenvalues 
in large-$N$ as
\[
\hbox{Mean number of real eigenvalues} \sim  \frac{2^{-\frac{N}{2}+2} \Gamma[N+1]}{\Gamma\left[\frac{N}{2}+1\right] \Gamma\left[\frac{N}{2}\right]} .
\]
In the large-$N$ limit, 
this expression indeed agrees with the result given 
in \cite{randommat, realnum2}.\footnote{{\color{blue}} Namely, Eq.(1.2) in \cite{realnum2}.} }

\begin{figure}
\begin{center}
\includegraphics[width=5cm]{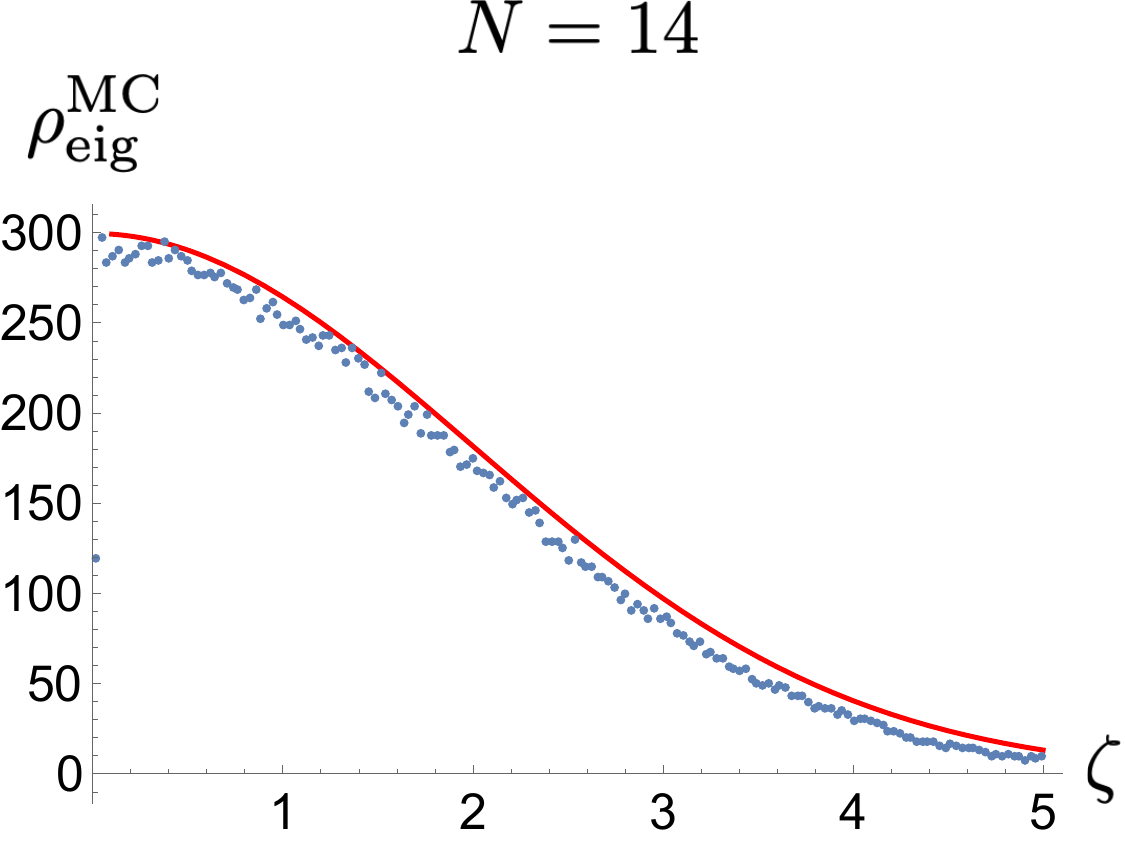}
\hfil
\includegraphics[width=5cm]{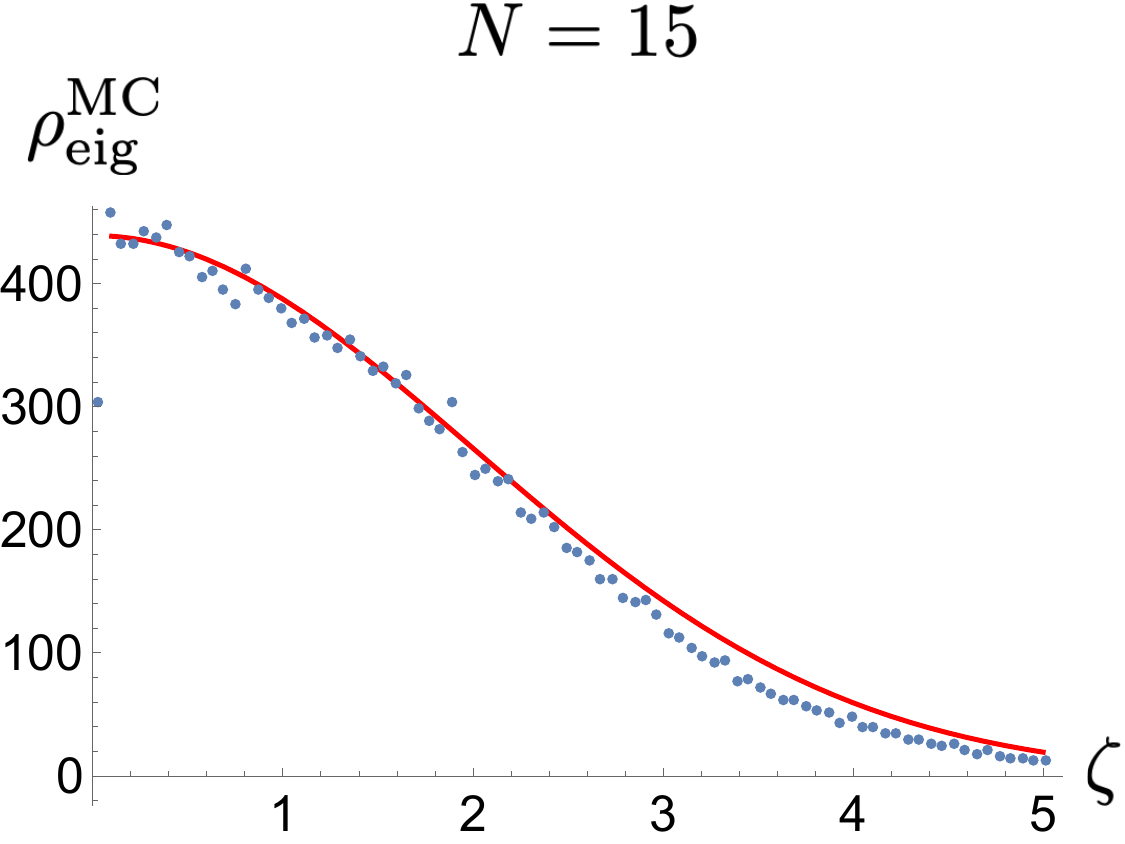}
\hfil
\includegraphics[width=5cm]{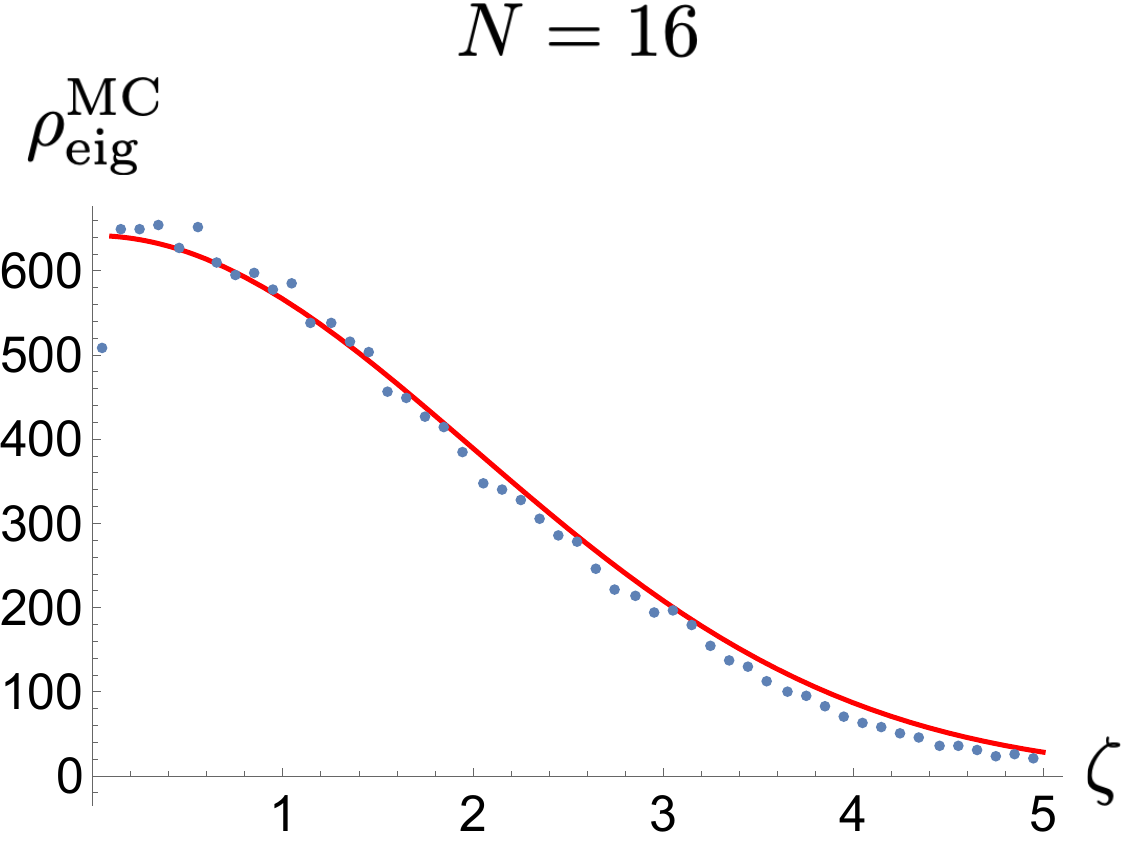}
\hfil
\caption{\label{fig:neq14} 
The guessed large-$N$ formula \eq{eq:app} (solid line) is compared 
with the Monte Carlo simulation \eq{eq:mcz} (dots). 
Left: $N=14, \delta \zeta=0.03, N_C=400$. Middle: $N=15, \delta \zeta=0.06, N_C=100$.
Right: $N=16, \delta \zeta=0.1, N_C=72$.}
\end{center}
\end{figure}

In Figure~\ref{fig:neq14} we compare the large-$N$ expression \eq{eq:app} with the Monte Carlo simulations 
for $N=14,15,16$.\footnote{With our machine power (See the last part of Section~\ref{subsec:odd}), it took one full day for the computation of the $N=16$ case. } 
They agree well, and the agreement seems to become slightly better, as $N$ becomes larger.

In the above discussion, we have not used the odd $N$ cases. The reason is that we could not find  
simple expressions valid across all the odd $N$ cases obtained in Section~\ref{sec:analytic}. 
We merely notice
\[
G_{N:{\rm odd}}=\pi^{N-1} z^\frac{N-1}{2} H_{N-1} \left( \frac{1}{2 \sqrt{z}}\right)
+d_0' \,e^{-\frac{1}{8z}} \left(1+\frac{d'_1}{z}+\cdots+\frac{d'_{N-3}}{z^{N-3}}\right),
\]
where we do not have any simple expressions for $d'_i$. 

\begin{figure}
\begin{center}
\hfil
\includegraphics[width=7cm]{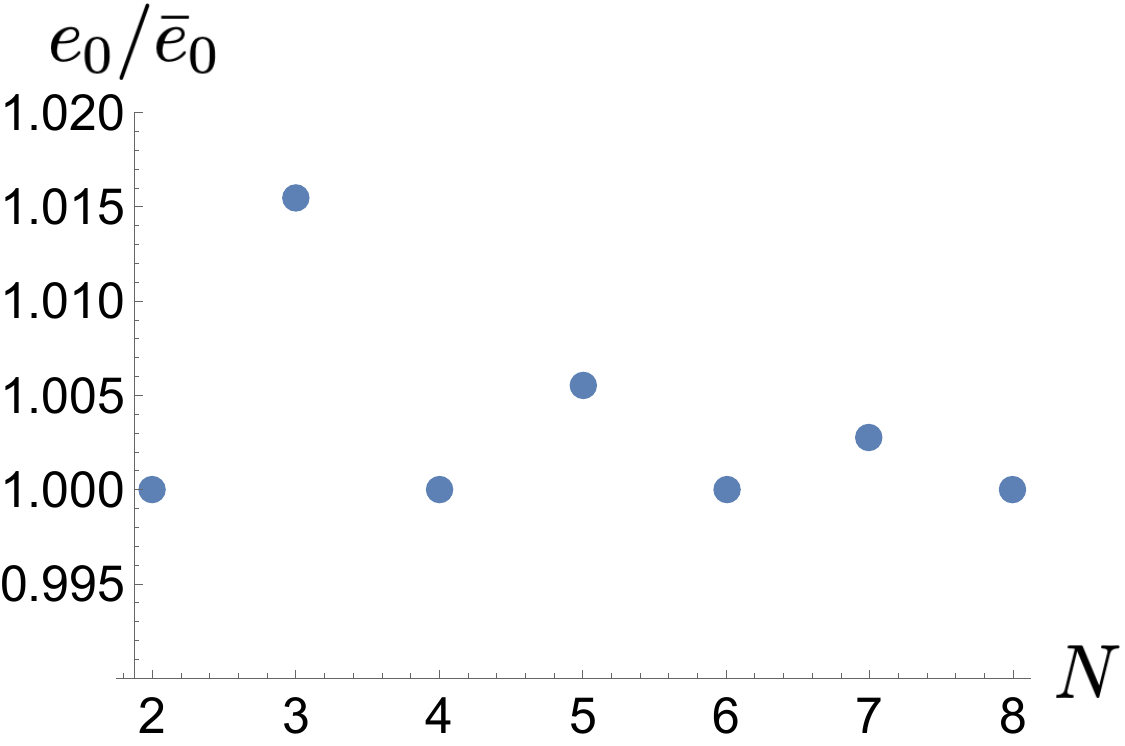}
\hfil
\includegraphics[width=7cm]{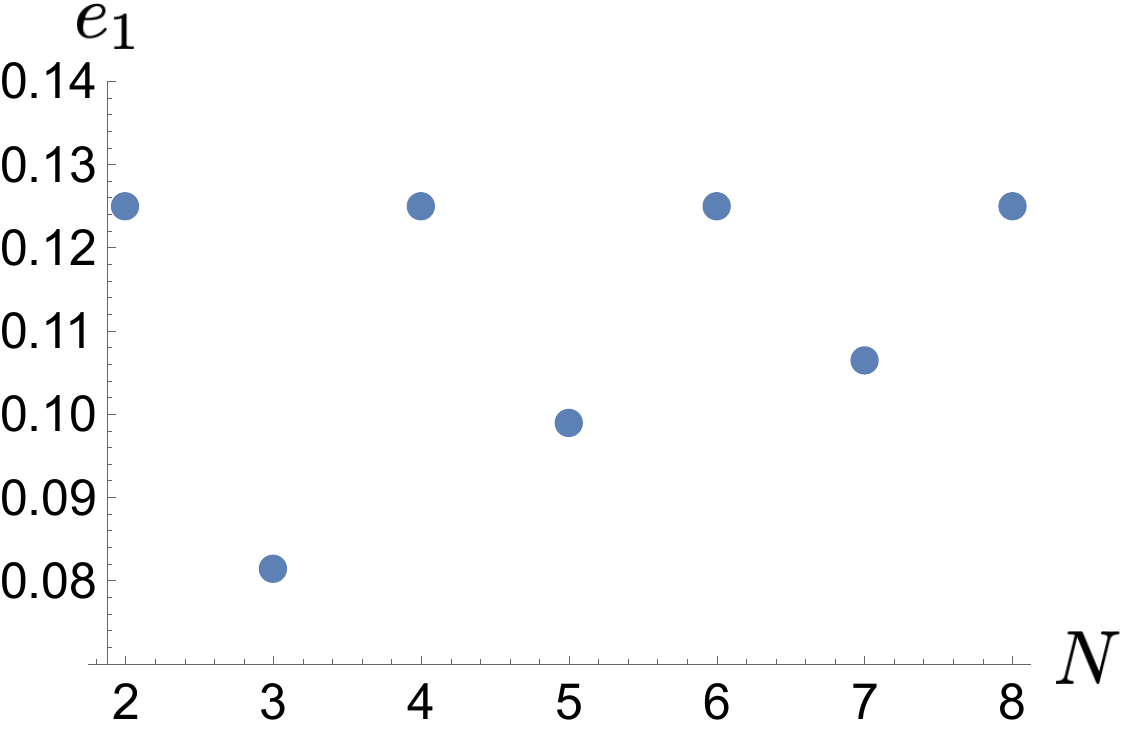}
\hfil
\caption{
\label{fig:e0e1}
The coefficients $e_0,e_1$ of the expansion \eq{eq:asymp} are plotted against $N$. 
$\bar e_0= \sqrt{2} \pi^{N-3/2} \Gamma[N]/\Gamma[N/2+1]$ (see \eq{eq:gnasymp}).
The values of $e_0,e_1$ for odd $N$ seem to approach those for even $N$, as $N$ becomes larger.
}
\end{center}
\end{figure}

Rather, we find that the large-$z$ behavior of $G_{N:{\rm odd}}$ approaches that of $G_{N:{\rm even}}$,
as $N$ becomes larger.  To see this, let us introduce the parameters of the $1/z$ expansion as
\[
G_N\sim e_0\, z^{\frac{N-1}{2}}\left(1+\frac{e_1}{z}+\cdots  \right).
\label{eq:asymp}
\]
As in \eq{eq:gnasymp}, for even $N$, we have 
\[
e_0=\frac{\sqrt{2} \pi^{N-\frac{3}{2}} \Gamma[N+1]}{\Gamma\left[\frac{N}{2}+1\right]},\ e_1=\frac{1}{8}.
\]
In Figure~\ref{fig:e0e1}, we plot $e_0,e_1$ for each $N$, which are extracted by performing the $1/z$ 
expansions of the explicit expressions in Section~\ref{sec:analytic}. 
Indeed $e_0,e_1$ for odd $N$ seem to approach those for even $N$, as $N$
becomes larger.
This implies that the large-$N$ formula \eq{eq:app} can commonly be used for both even and odd $N$.

\section{Summary and future prospects}
In this paper we have explicitly computed the real eigenvalue/vector distributions for
real symmetric order-three tensors with Gaussian distributions for $N\leq 8$. 
This has been achieved by rewriting the problem as the computation of a partition 
function of a zero-dimensional boson-fermion system with four-interactions. We have extrapolated the
exact expressions for $N\leq 8$ to guess a large-$N$ expression.
Monte Carlo simulations have been performed, and precise agreement and good agreement 
have been obtained for the 
exact small-$N$ expressions and the large-$N$ expression, respectively.

It seems surprising that the complicated integrations of the partition function can be performed exactly with
the final simple results expressed by polynomial, exponential, and error functions. 
It is suspected that this is true for any $N$, but deeper understanding of this aspect needed for 
such generalization is left for future study. 
Considering the final simple forms, it would be possible that there exist much easier
ways to compute the distributions than what has been done in this paper. 

There would exist various directions to pursue by a similar strategy of this paper. 
One obvious direction is the generalization of tensors to those with different orders 
and non-Gaussian distributions.
It would also be interesting to pursue the possibility of 
topological changes of the distributions in tensor models as in matrix models.
In fact, some tensor systems seemingly suggest such possibilities \cite{Kawano:2021vqc,Sasakura:2022aru}.
It would also be possible to compute correlations among eigenvalues/vectors, {{\color{blue}} 
and also to incorporate complex eigenvalues/vectors.}

\vspace{.3cm}
\section*{Acknowledgements}
The author is supported in part by JSPS KAKENHI Grant No.19K03825. 

\appendix
\def\thesection{Appendix  \Alph{section}}
\counterwithin{equation}{section}
\renewcommand{\theequation}{\Alph{section}\arabic{equation}}

\section{Computations of $E_2$ and $E_3$}
\label{app:e2e3}
From \eq{eq:delscterms},
\[
E_2=-\frac{4}{\alpha}A_1,
\]
with 
\[
A_1=\left( \frac{1}{6} \sum_s v_{s_a} \sigma_{s_b} \phi_{s_c} \right)^2.
\]
By explicitly writing down the sum and noting that one of the symmetrization is not necessary, 
we obtain
\s[
A_1&=\frac{1}{6} v_a \sigma_b \phi_c (v_a \sigma_b \phi_c+v_b \sigma_c \phi_a+v_c \sigma_a \phi_b
+v_b \sigma_a \phi_c+v_a \sigma_c \phi_b+v_c \sigma_b \phi_a) \\
&=\frac{1}{6}
\left( 
v^2 \sigma^2 \phi^2+v\cdot \sigma\, \sigma\cdot \phi\, \phi\cdot v+v\cdot \phi\, \sigma\cdot v \, \phi\cdot \sigma+
v\cdot \sigma\, \sigma\cdot v \, \phi^2+v^2\, \sigma\cdot \phi\, \phi\cdot \sigma+v\cdot \phi\,\sigma^2\,\phi\cdot v
\right) \\
&=\frac{|v|^2}{6} \left(
\sigma^2 \phi^2+2 \pphi\psig\,\sigma\cdot \phi+\psig^2\phi^2+\pphi^2\sigma^2+(\sigma\cdot\phi)^2
\right)\\
&=\frac{|v|^2}{6}\left(
\tsig^2 \tphi^2+(\tphi\cdot \tsig)^2+2 \psig^2 \tphi^2+2 \tsig^2 \pphi^2+4 \psig \pphi \, \tphi\cdot\tsig+6 \psig^2 \pphi^2 
\right),
\s]
where we have used $v\cdot \phi=|v| \pphi$, etc., and $\phi\cdot\sigma=\pphi\psig+\tphi\cdot \tsig$, etc..

Let us next compute 
\[
E_3 = \frac{4 i}{\alpha} A_2,
\]
where 
\[
A_2&=(v_{a} \bpsi_{b} \vphi_{c} +v_{a} \bvphi_{b}\psi_{c} )\, 
\frac{1}{6} \sum_s v_{s_a} \sigma_{s_b} \phi_{s_c},
\]
from \eq{eq:delscterms}. By explicitly expanding the sum, we obtain
\s[
A_2
&= v_a \bpsi_b \vphi_c (v_a \sigma_b \phi_c+v_b\sigma_c \phi_a+v_c \sigma_a \phi_b+v_b \sigma_a \phi_c+v_a \sigma_c \phi_b+v_c \sigma_b \phi_a)+(\psi\leftrightarrow \vphi) \\
&=\frac{|v|^2}{6} \big(
\tbpsi\cdot \tsig\,\tvphi\cdot \tphi+\tbpsi\cdot\tphi\,\tvphi\cdot\tsig+\tbvphi\cdot \tsig\,\tpsi\cdot\tphi
+\tbvphi\cdot\tphi\,\tpsi\cdot\tsig  \\
&\ \ \ \ \ \ \ \ \ +2 \pbpsi \psig \tvphi\cdot \tphi+2 \pphi \pbpsi \tvphi\cdot \tsig+
2\pvphi \pphi \tbpsi\cdot \tsig +2 \pvphi\psig \tbpsi\cdot \tphi \\
&\ \ \ \ \ \ \ \ \ +2 \pbvphi \psig \tpsi\cdot\tphi+2 \pphi \pbvphi \tpsi\cdot \tsig+2 \ppsi \pphi \tbvphi\cdot \tsig+2 \ppsi \psig \tbvphi\cdot \tphi\\
&\ \ \ \ \ \ \ \ \ +6 \pbpsi \pvphi \pphi \psig+6 \pbvphi \ppsi \pphi \psig
\big),
\s]
where we have taken similar computational steps as those for $A_1$.

\section{Lists of $a_i$ and $b_i$ for $N=5,7$}
\label{app:odd}
For $N=5$, 
\s[
a_0 &= 1 - 8 z + 120 z^2 + 480 z^3 + 2640 z^4, \\
a_1 &= -16 i (z - 6 z^2 + 60 z^3 + 120 z^4),  \\
a_2 &= -32 (3 z^2 - 10 z^3 + 100 z^4 + 200 z^5), \\
a_3 &=  -32 (-z^2 + 10 z^3 + 20 z^4 + 280 z^5), \\
a_4 &= -256 i (z^3 + 20 z^5), \\
a_5 &= 256 (z^4 + 4 z^5 + 28 z^6),
\s]
and 
\s[
b_0&=  -4 (1 - 12 z + 12 z^2), \\
b_1&=  4 (-1 + 22 z - 132 z^2 + 120 z^3), \\
b_2&=   48 (z - 8 z^2 + 60 z^3 + 80 z^4), \\
b_3&=   -128 (z^2 + 60 z^4 + 240 z^5).
\s]
For $N=7$,
\s[
a_0 &= 1 - 36 z + 660 z^2 - 3360 z^3 + 25200 z^4 + 100800 z^5 + 
   504000 z^6, \\
a_1 &=  -16 i (z - 30 z^2 + 440 z^3 - 1680 z^4 + 8400 z^5 + 16800 z^6),\\
a_2 &= -32 (3 z^2 - 70 z^3 + 840 z^4 - 1680 z^5 + 11760 z^6 + 
     30240 z^7), \\
a_3 &= -32 (-z^2 + 30 z^3 - 280 z^4 + 1680 z^5 + 5040 z^6 + 
     36960 z^7), \\
a_4 &= -256 i (z^3 - 16 z^4 + 168 z^5 + 1680 z^7), \\
a_5 &= 256 (z^4 - 8 z^5 + 120 z^6 + 480 z^7 + 2640 z^8),
\s]
and
\s[
b_0&= 8 (-1 + 30 z - 180 z^2 + 120 z^3), 
\\ b_1&= -8 (1 - 44 z + 600 z^2 - 2640 z^3 + 1680 z^4), 
\\ b_2&= 4 (-1 + 58 z - 1160 z^2 + 9360 z^3 - 28560 z^4 + 16800 z^5), 
\\ b_3&= 16 (3 z - 100 z^2 + 1480 z^3 - 6720 z^4 + 25200 z^5 + 
    33600 z^6), 
\\ b_4&= -128 (z^2 - 20 z^3 + 280 z^4 + 8400 z^6 + 33600 z^7).
\s]

\section{Lists of $a_i,b_i,c_0$ for $N=4,6,8$}
\label{app:even}
For $N=4$,
\s[
a_0 &= 1 + 60 z^2 + 240 z^3, \\
a_1 &= -16 i (z + 20 z^3), \\
a_2 &=  -32 (3 z^2 + 2 z^3 + 24 z^4), \\
a_3 &= -32 (-z^2 + 6 z^3 + 32 z^4),\\
a_4 &=   -256 i (z^3 + 2 z^4), \\
a_5 &=  256 (z^4 + 4 z^5), 
\s]
and 
\s[
c_0&= -2 (-1 + 6 z), \\
b_0 &= 4 (-1 + 6 z), \\ 
b_1 &=16 (3 z - 2 z^2 + 40 z^3), \\
b_2 &=  -128 (z^2 + 4 z^3 + 28 z^4).
\s]
For $N=6$, 
\s[
a_0 &= 1 - 20 z + 280 z^2 + 8400 z^4 + 33600 z^5, \\
a_1 &= -16 i (z - 16 z^2 + 168 z^3 + 1680 z^5), \\
a_2 &= -32 (3 z^2 - 34 z^3 + 300 z^4 + 360 z^5 + 2400 z^6), \\
a_3 &=  -32 (-z^2 + 18 z^3 - 60 z^4 + 600 z^5 + 2880 z^6), \\
a_4 &= -256 i (z^3 - 6 z^4 + 60 z^5 + 120 z^6), \\
a_5 &= 256 (z^4 + 60 z^6 + 240 z^7), 
\s]
and 
\s[
c_0&=  3 (1 - 20 z + 60 z^2), \\
b_0&= -6 (1 - 20 z + 60 z^2), \\
b_1&= 4 (-1 + 18 z) (1 - 20 z + 60 z^2),  \\
b_2 &= 16 (3 z - 56 z^2 + 540 z^3 - 240 z^4 + 3360 z^5), \\
b_3 &= -128 (z^2 - 8 z^3 + 120 z^4 + 480 z^5 + 2640 z^6).
\s]
For $N=8$, 
\s[
a_0 &=  1 - 56 z + 1428 z^2 - 15120 z^3 + 105840 z^4 + 2116800 z^6 + 
  8467200 z^7,\\
a_1 &= -16 i (z - 48 z^2 + 1020 z^3 - 8640 z^4 + 
    45360 z^5 + 302400 z^7), \\
a_2 &= -32 (3 z^2 - 118 z^3 + 2080 z^4 - 
    12880 z^5 + 58800 z^6 + 84000 z^7 + 
    470400 z^8), \\
a_3 &= -32 (-z^2 + 46 z^3 - 760 z^4 + 6160 z^5 - 
    8400 z^6 + 117600 z^7 + 537600 z^8), \\
a_4 &= -256 i (z^3 - 30 z^4 + 
    440 z^5 - 1680 z^6 + 8400 z^7 + 16800 z^8), \\
a_5 &= 
 256 (z^4 - 20 z^5 + 280 z^6 + 8400 z^8 + 33600 z^9),
\s]
and 
\s[
c_0&= -(15/2) (-1 + 42 z - 420 z^2 + 840 z^3), \\
b_0&= 15 (-1 + 42 z - 420 z^2 + 840 z^3), \\
b_1&= -10 (-1 + 24 z) (-1 + 42 z - 420 z^2 + 840 z^3), \\
b_2 &= 4 (1 - 40 z + 360 z^2) (-1 + 42 z - 420 z^2 + 840 z^3), \\
b_3 &= 48 (z - 52 z^2 + 1140 z^3 - 10360 z^4 + 50400 z^5 - 
    16800 z^6 + 201600 z^7), \\
b_4 &= -128 (z^2 - 36 z^3 + 660 z^4 - 3360 z^5 + 25200 z^6 + 
    100800 z^7 + 504000 z^8).
\s]

{{\color{blue}}
\section{Complexity of the spherical $p$-spin model}
\label{app:relation}
Computing the distributions of the real tensor eigenvalues/vectors is the same as the computation of 
complexity\footnote{{\color{blue}} See for example \cite{example} for a review.} 
of the spherical $p$-spin model \cite{pspin,pedestrians}. 
In this appendix, we limit ourselves to $p=3$ corresponding to order-three tensors.

The Hamiltonian of the spherical $p$-spin model (with $p=3$) is given by 
\[
H=-\frac{1}{N} J_{abc} \sigma_a \sigma_b \sigma_c, 
\label{eq:hamiltonian}
\]
with the continuous spin variable $\sigma \in \mathbb{R}^N$ which has a constraint, 
\[
\sigma_a \sigma_a =N.
\label{eq:normalsigma}
\]
Here the real symmetric tensor $J$ is assumed to have the normal Gaussian distribution 
and can be identified with $C$ with $\alpha=1/2$ in our notation. 
To match \eq{eq:normalsigma} with our notation, we introduce the change of variable,
\[
\sigma_a=\sqrt{N} w_a,
\]
and then we have
\[
H=-\sqrt{N} C_{abc} w_a w_b w_c \hbox{ with } w_a w_a=1.
\label{eq:hampspin}
\]

The problem of computing complexity of the $p$-spin spherical model is 
to count the number of local minima (and stationary points) of the 
above Hamiltonian.
By using the method of Lagrange multiplier, counting the stationary points is equivalent to solve 
\[
C_{abc} w_b w_c=-\sqrt{N} u w_a,
\label{eq:egham}
\]
where $u\in \mathbb{R}$ and $w_a w_a=1 \ (w\in \mathbb{R}^N)$.
The value of $u$ is the energy of the stationary point, and can be identified with the 
same variable employed in \cite{randommat}.
Comparing with the Z-eigenvalue equation \eq{eq:zegeq}, we have the relation,
\[
\zeta=-\sqrt{N} u.
\label{eq:zu}
\]
By using \eq{eq:zu}, it can be checked that the expressions of the mean distributions of the numbers of
the critical points given for general $N$ in Section 7.2 of \cite{randommat} agree with our results in Section~\ref{sec:analytic}.
}

\end{document}